\documentclass[a4paper,11pt]{article}
\usepackage{pos}
\usepackage[capitalise]{cleveref}
\usepackage{subcaption}
\usepackage{siunitx}

\newcommand{\DD}{\mathrm{D}}

\newcommand{\Dz}{\mathrm{D}^{0}}
\newcommand{\Dzbar}{\overline{\mathrm{D}^{0}}}

\title{Towards new D meson fragmentation functions}

\author[a]{Manuel Epele}
\author*[b,c]{Felix Hekhorn}
\author[b,c]{Ilkka Helenius}
\author[b,c]{Hannu Paukkunen}
\author[d]{Pia Zurita}

\affiliation[a]{IFLP, CONICET - Dpto. de F\'isica, Universidad Nacional de La Plata, C.C. 67, 1900 La Plata, Argentina}
\affiliation[b]{University of Jyvaskyla, Department of Physics, P.O. Box 35, FI-40014 University of Jyvaskyla, Finland}
\affiliation[c]{Helsinki Institute of Physics, P.O. Box 64, FI-00014 University of Helsinki, Finland}
\affiliation[d]{Departamento de F\'isica Te\'orica \& IPARCOS, Universidad Complutense de Madrid, Plaza de las Ciencias 1, E-28040 Madrid, Spain}

\emailAdd{felix.a.hekhorn@jyu.fi}
\abstract{
  The Heavy Meson (Hymn) collaboration presents a new extraction of D meson fragmentation functions using experimental data from LEP and LHC.
  We focus particularly on kinematical regimes where perturbative QCD should be safely applicable to avoid contamination from higher-twist effects which could lead to an apparent process dependence of fragmentation functions.
  We account for the initial-state radiation and, as a novel ingredient, consider the prompt and non-prompt contributions separately.
  The analysis is carried out at next-to-leading order accuracy including uncertainty estimation based on Monte-Carlo replica technique.
  We disucss the exemplary case of $\Dz$ here and defer the results for $\DD^\pm$ and $\DD^{*,\pm}$ to a forthcoming publication.
}

\FullConference{The 33rd International Workshop on Deep Inelastic Scattering and Related Subjects (DIS2026)\\
4 - 8 May 2026\\
Bologna, Italy\\}

\begin{document}
\maketitle

\section{Introduction}
The fragmentation of heavy mesons, such as D mesons~\cite{DFF}, has received renewed attention in recent years.
On the one side experimental measurements taken at hadron colliders, such as the Large Hadron Collider (LHC), and at lepton-lepton colliders, such as the Large Electron-Positron Collider (LEP), or lepton-hadron colliders show some tension, in particular, for the ratio of produced charmed mesons to charmed baryons.
We refer the reader to Ref.~\cite{dEnterria:2026tuz} for a detailed review on the topic.
On the other side, fragmentation functions (FFs) $D_{j/\DD}(z,\mu^2)$ are defined as universal objects in the framework of collinear factorization as follows from their definition in terms of quantum field operators~\cite{Collins:1989gx}, where $j$ refers to the parton flavor, $z$ to its momentum fraction, and $\mu^2$ to the fragmentation scale.
Specifically, the factorization formula for Single Inclusive Annihilation (SIA),
\begin{equation}
    \sigma^{e^+ + e^- \to \DD + X}(q^2) = c(q^2) \otimes D_{\DD}(q^2) + \mathcal O(1/q^2)\,, \label{eq:SIAfact}
\end{equation}
with $q^2$ referring to the virtuality of the electro-weak boson, and the factorization formula for hadro-production of D mesons,
\begin{equation}
    \sigma^{p + p \to \DD + X}(p_T^2) = f(p_T^2) \otimes f(p_T^2) \otimes \hat \sigma(p_T^2) \otimes D_{\DD}(p_T^2) + \mathcal O(1/p_T)\,, \label{eq:Hadrofact}
\end{equation}
with $p_T$ referring to the transverse momentum of the meson, contain the same, universal FF $D_\DD$.
The coefficient functions $c$ in \cref{eq:SIAfact} and the partonic matrix elements $\hat \sigma$ in \cref{eq:Hadrofact} are directly computable in perturbative Quantum Chromo Dynamics (QCD), while the parton distribution functions (PDFs) $f$ share similar universality properties as FFs.

We note that the sub-leading terms in \cref{eq:SIAfact} and \cref{eq:Hadrofact} are process-dependent and may thus provide a possible explanation of the experimentally observed tension.
Our aim here is to obtain a new set of D meson FFs $D_\DD$, which are safe from such potential contaminations and which improve the currently available extractions in other areas at the same time.
In our forthcoming publication~\cite{DFF} we will investigate the case of three meson species, $\Dz$, $\DD^\pm$, and $\DD^{*,\pm}$, but discuss in this work only the exemplary case of $\Dz$.

\section[Fitting D0 fragmentation functions]{Fitting $D^0$ fragmentation functions}
\begin{figure}
    \centering
    \begin{subcaptionblock}{.33\textwidth}
        \includegraphics[width=\textwidth]{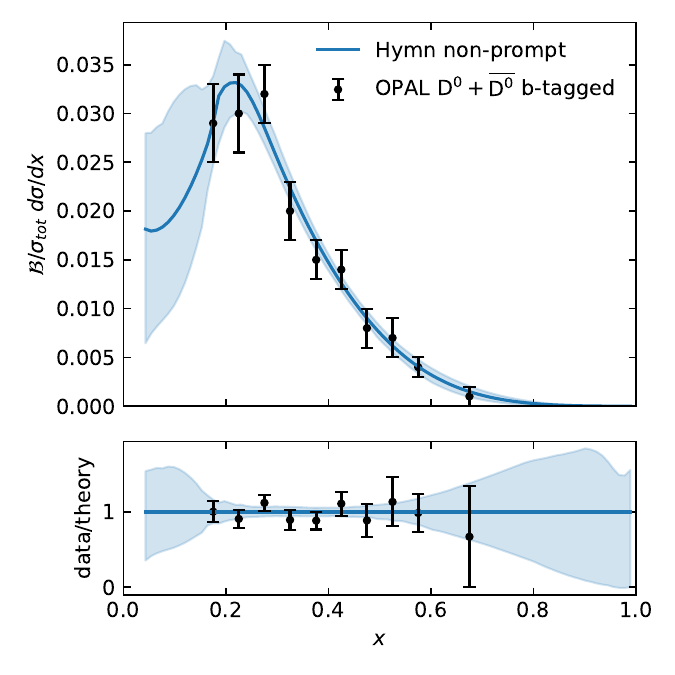}
        \caption{non-prompt SIA}
        \label{fig:np}
    \end{subcaptionblock}%
    \begin{subcaptionblock}{.33\textwidth}
        \includegraphics[width=\textwidth]{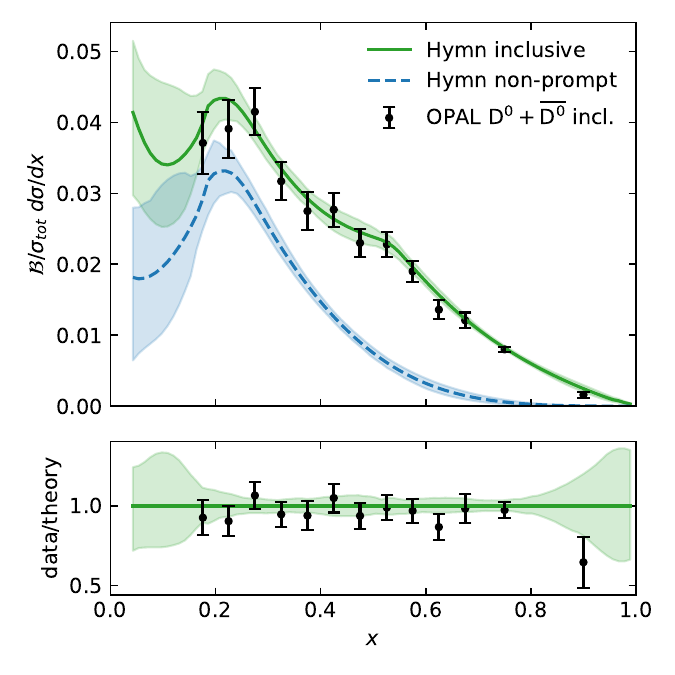}
        \caption{inclusive SIA}
        \label{fig:inc}
    \end{subcaptionblock}%
    \begin{subcaptionblock}{.33\textwidth}
        \includegraphics[width=\textwidth]{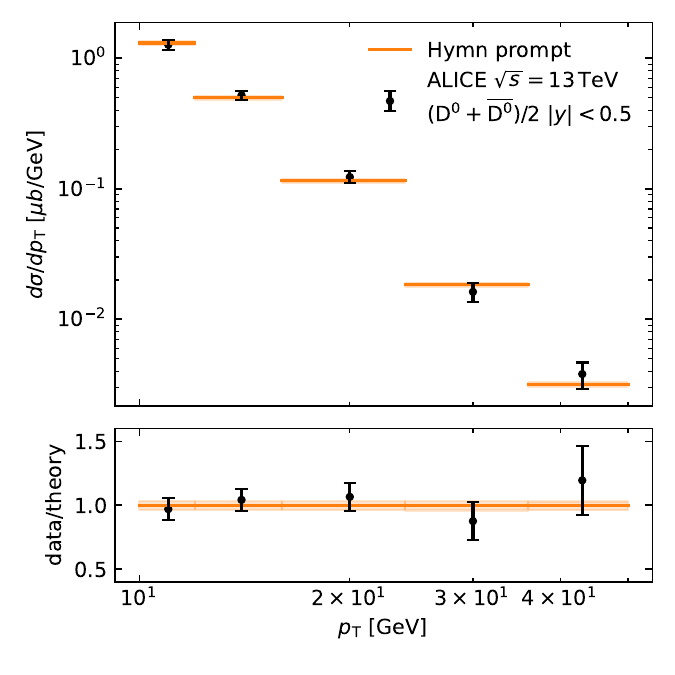}
        \caption{hadro-production}
        \label{fig:hp}
    \end{subcaptionblock}
    \caption{Exemplary comparison between experimental data from OPAL~\cite{OPAL:1996ikk} and ALICE~\cite{ALICE:2023sgl} to theory predictions for $\Dz$.
    }
    \label{fig:data}
\end{figure}

We extract the FFs by performing a fit to experimental measurements using the xFitter framework~\cite{Alekhin:2014irh,Zurita:2021kli}.
We consider both SIA measurements (from DELPHI~\cite{DELPHI:1993gqe} and OPAL~\cite{OPAL:1996ikk}) and hadro-production measurements (from ALICE~\cite{ALICE:2023sgl} and CMS~\cite{CMS:2021lab}).
For the latter we impose a cut on minimum $p_{\mathrm{T}}$ of the D meson of $\SI{10}{\GeV}$.
For our FFs we explicitly distinguish \textit{non-prompt} and \textit{prompt} production mechanisms.
In the former, the D mesons are decay products of B hadrons whereas in the latter the D mesons are produced from charm-quark fragmentation.
Both processes obey their own factorization formula, and so does the sum of these two contributions to which we refer to as the \textit{inclusive} production.
Indeed, we have separate SIA measurements for non-prompt and inclusive measurements, while, instead, the hadro-production measurements refer only to the prompt component.
\Cref{fig:data} shows a representative subset of these different measurements compared with the Hymn analysis~\cite{DFF}.

In order to disentangle the two independent components we employ a multi-stage fitting procedure:
first, we extract the non-prompt distribution;
second, using this FF, we compute the respective contribution to the inclusive measurements;
third, we extract the prompt distribution.
For the non-prompt component the fitted non-perturbative input is the FF $D_{b/\Dz}^\text{np}(z,m_b^2)$ at the bottom quark mass $m_b$ and for the prompt component the FF $D_{c/\Dz}^\text{pr}(z,m_c^2)$ at the charm quark mass $m_c$.
Both input distributions are parametrized with an analytic function with four free parameters each.
The FF uncertainties are quantified via the Monte-Carlo framework~\cite{Costantini:2024wby} by generating independent data replicas and performing the above sequence of step individually in each case.
The resulting set of FFs is added to our final ensemble provided it passes our acceptance criteria, such as a faithful fit convergence.
We use next-to-leading order (NLO) perturbation theory and account for additional initial-state photon radiation in SIA~\cite{Kneesch:2007ey}.
Since we only include measurements with a large scale we can safely neglect any mass corrections.
We explicitly distinguish between the meson and its anti-meson, i.e.\ in the case here between $\Dz$ and $\Dzbar$, and we relate them by charge conjugation.

\section[First results for D0]{First results for $\Dz$}
We now turn to the extracted FFs themselves.
For the non-prompt fit we have ten data points (see \cref{fig:np}), which yield a goodness-of-fit $\chi^2_\text{tot}/N_\text{dof} = 0.76$.
For the prompt fit we have 20 inclusive SIA data points (a subset is shown in \cref{fig:inc}) and ten prompt hadro-production data points (a subset is shown in \cref{fig:hp}), which all together yield a goodness-of-fit $\chi^2_\text{tot}/N_\text{dof} = 1.07$.

\begin{figure}
  \includegraphics[width=\textwidth]{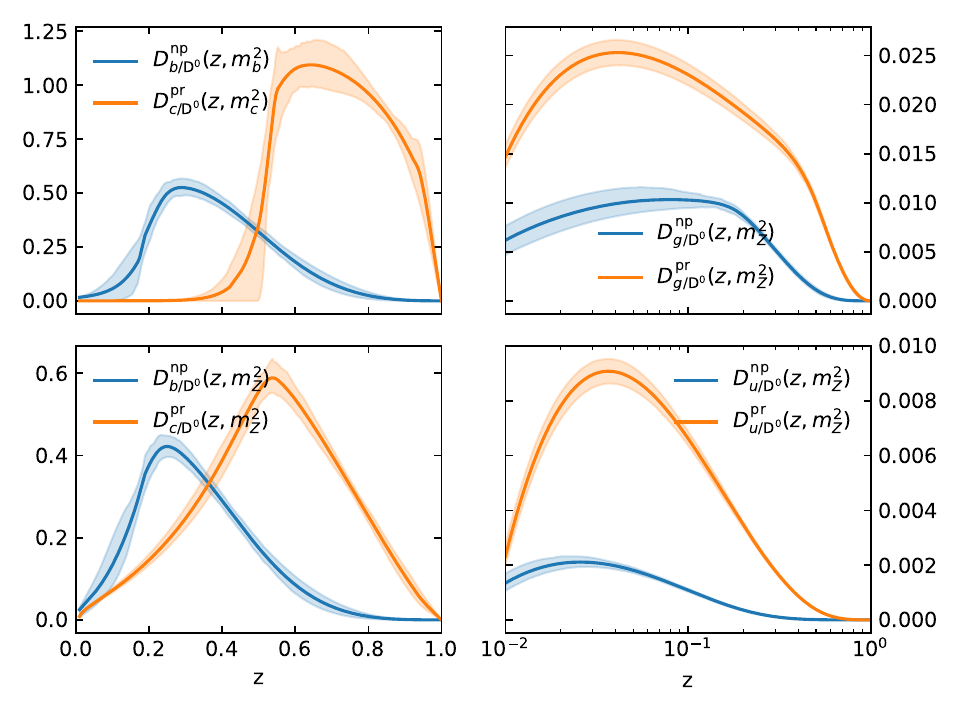}
  \caption{$\Dz$ non-prompt (blue) and prompt (orange) FFs.
    The left column shows the fitted distributions, i.e.\ bottom distribution for non-prompt and charm distribution for prompt, at the parametrization scale (top) and the scale $\mu=m_Z$ (bottom).
    The right column shows two evolution-driven distributions, the gluon distribution (top) and the up distribution (bottom), at the scale $\mu=m_Z$.
    The shaded error bands represent the 68~\% confidence interval.}
  \label{fig:FF}
\end{figure}
\Cref{fig:FF} gives an overview of our fitted distributions.
We always plot the non-prompt and prompt FFs in the same panel using various flavors and fragmentation scales $\mu$.
We observe that the two components obey the expected hierarchy:
first, the prompt distribution mostly dominates over the non-prompt counterpart, and, second, the non-prompt distribution peaks at smaller momentum fractions $z$ as is expected since they fragment into heavier mesons first.
Having distinguished the two components allows us to disentangle for the first time their respective gluon components, which are purely generated via evolution, and we find that the non-prompt gluon can make up to 40~\% of its prompt counterpart.

\begin{figure}
    \centering
    \includegraphics[width=0.95\linewidth]{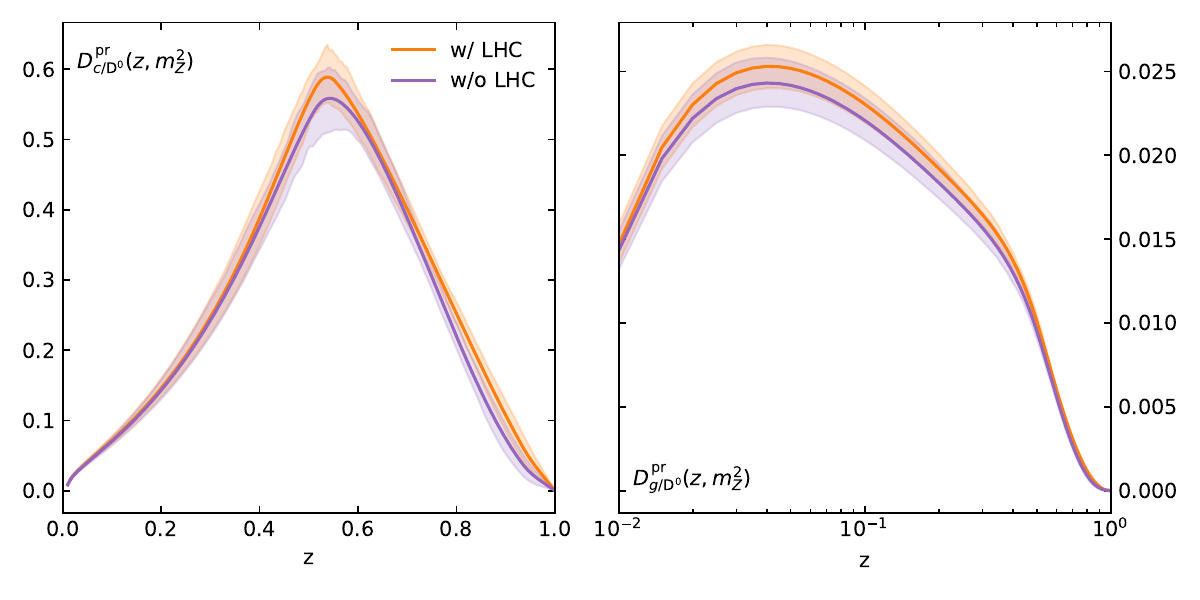}
    \caption{Comparison of prompt FFs obtained from fitting with (orange) and without (purple) LHC data.
    The left (right) panels show the charm (gluon) distribution as a function of the momentum fraction $z$ at the scale $\mu = m_Z$.}
    \label{fig:LHC}
\end{figure}
While the LHC data is included in our default extraction, we determine its impact explicitly in a dedicated fit where we exclude it and which is shown in \cref{fig:LHC}.
We recall that the LHC measurements put a more direct constraint on the gluon distribution in contrast to SIA measurements, where it only enters through NLO contributions and evolution effects.
We find that the LHC data favors slightly higher charm and gluon distributions, but the two determinations are compatible within their statistical uncertainty.

\begin{figure}
  \includegraphics[width=\textwidth]{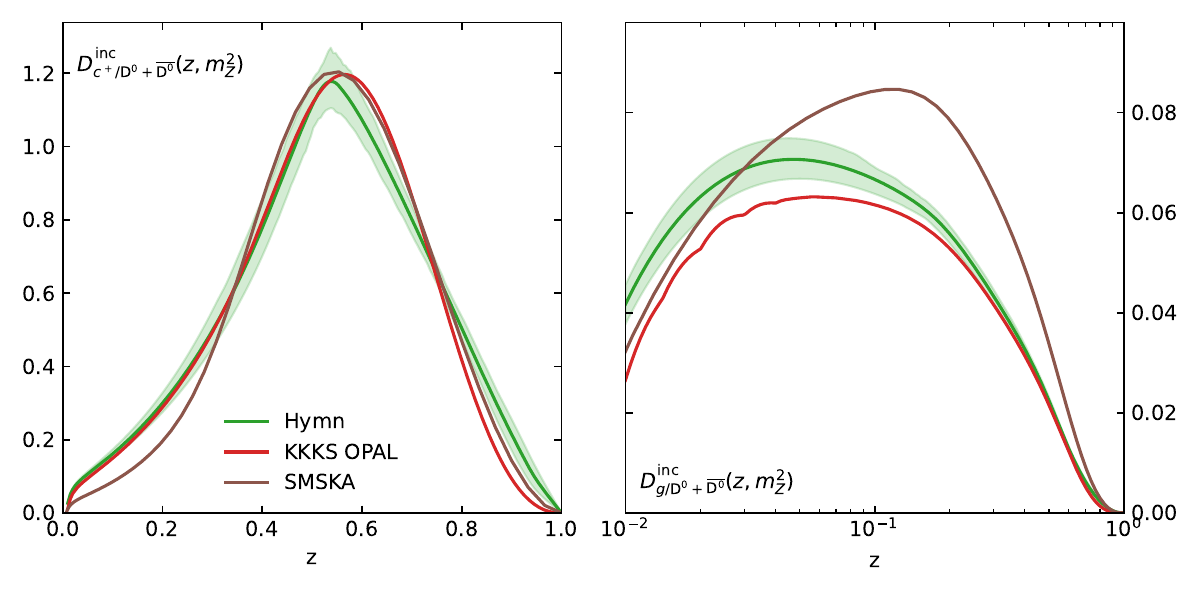}
  \caption{Comparison of inclusive FFs obtained from us (green) against KKKS~\cite{Kneesch:2007ey} using OPAL data, and SMSKA~\cite{Salajegheh:2019nea} (brown).
    The left (right) panel shows the total charm (gluon) distribution as a function of the momentum fraction $z$ at the scale $\mu = m_Z$.}
  \label{fig:comp}
\end{figure}
We compare in \cref{fig:comp} our extractions, denoted by \textit{Hymn}, against earlier determinations by KKKS~\cite{Kneesch:2007ey} (OPAL variant) and SMSKA~\cite{Salajegheh:2019nea}.
We plot the inclusive total charm and gluon distribution at the scale $\mu = m_Z$.
We find overall very good agreement for the case of charm, where the results agree within the provided error estimate.
While the agreement is slightly worse in the case of gluon, we recall that we consider non-prompt and prompt separately unlike earlier determinations, which may cause some mismatch.
Finally, we recall that FFs are basically unconstrained by data in the small $z$-region and all extractions are instead subject to parametrization biases.

\section{Conclusion and outlook}
Our forthcoming publication~\cite{DFF} will present new sets of FFs for D mesons, specifically, $\Dz$, $\DD^\pm$, and $\DD^{*,\pm}$, out of which we have highlighted the $\Dz$ results here.
We update and improve earlier determinations in several aspects.
We distinguish, for the first time, between non-prompt, prompt, and inclusive production mechanisms and provide separate distributions for each component.
Moreover, we distinguish, for the first time, between the meson (e.g.\ $\Dz$), its anti-meson (e.g.\ $\Dzbar$) and their sum (e.g.\ $\Dz + \Dzbar$).
Eventually, this yields nine distributions per meson, enabling user to pick a suitable one corresponding to the studied case.

We use measurements from LEP (via SIA) and the LHC (via hadro-production) to extract our FFs.
We restrict our data selection such that we are less susceptible to higher-twist or power corrections, which could potentially spoil the universality of FFs.
This will serve as a potential baseline to study the experimentally observed tension between charmed meson and baryon formation.
We provide our fitting error estimates using MC replicas via the LHAPDF6 interface~\cite{Buckley:2014ana}, which will facilitate the use of our results by the community.

In the future we plan to extend our framework to other hadrons, use more data, and, eventually, address the limit of collinear factorization directly.

\acknowledgments

We acknowledge grants of computer capacity from the Finnish Grid and Cloud Infrastructure (persistent identifier urn:nbn:fi:research-infras-2016072533).
M.E.'s work has been partially supported by CONICET and Agencia I+D+i via FONCyT project 01-PICT 2022-2022-11-00204.
M.E.\ acknowledges the support of the University of Jyväskylä during his postdoctoral fellowship.
F.H.\ and I.H.\ have been supported by the Research Council of Finland project 361179.
F.H., I.H., and H.P.\ were funded as a part of the Center of Excellence in Quark Matter of the Research Council of Finland, project 364194.
P.Z.\ is funded by the \textit{Atracci\'on de Talento Investigador} program of the Comunidad de Madrid (Spain) No.~2022-T1/TIC-24024.
P.Z.\ acknowledges the support of the research visitor program of the University of Jyv\"askyl\"a where part of this work was conducted.

\bibliographystyle{utphys}
\bibliography{refs.bib}

\providecommand{\href}[2]{#2}\begingroup\raggedright\begin{thebibliography}{10}

\bibitem{DFF}
M.~Epele, F.~Hekhorn, I.~Helenius, H.~Paukkunen, and P.~Zurita, ``{Hymn D v1: New D meson fragmentation functions from LEP and LHC},'' in preparation, 2026.

\bibitem{dEnterria:2026tuz}
D.~d'Enterria, F.~Hekhorn, I.~Helenius, V.~D. Le, and H.~Paukkunen, ``{Inclusive charm and bottom quark pair production cross sections at hadron colliders at next-to-next-to-leading-order accuracy},'' \href{https://arxiv.org/abs/2605.16019}{{\ttfamily arXiv:2605.16019 [hep-ph]}}.

\bibitem{Collins:1989gx}
J.~C. Collins, D.~E. Soper, and G.~F. Sterman, ``{Factorization of Hard Processes in QCD},'' \href{https://dx.doi.org/10.1142/9789814503266_0001}{{\em Adv. Ser. Direct. High Energy Phys.} {\bfseries 5} (1989) 1--91}, \href{https://arxiv.org/abs/hep-ph/0409313}{{\ttfamily arXiv:hep-ph/0409313}}.

\bibitem{Alekhin:2014irh}
S.~Alekhin {\em et~al.}, ``{HERAFitter},'' \href{https://dx.doi.org/10.1140/epjc/s10052-015-3480-z}{{\em Eur. Phys. J. C} {\bfseries 75} no.~7, (2015) 304}, \href{https://arxiv.org/abs/1410.4412}{{\ttfamily arXiv:1410.4412 [hep-ph]}}.

\bibitem{Zurita:2021kli}
P.~Zurita, ``{Medium modified Fragmentation Functions with open source xFitter},'' \href{https://arxiv.org/abs/2101.01088}{{\ttfamily arXiv:2101.01088 [hep-ph]}}.

\bibitem{DELPHI:1993gqe}
{\bfseries DELPHI} Collaboration, P.~Abreu {\em et~al.}, ``{A Measurement of D meson production in Z0 hadronic decays},'' \href{https://dx.doi.org/10.1007/BF01562545}{{\em Z. Phys. C} {\bfseries 59} (1993) 533--546}. [Erratum: Z.Phys.C 65, 709--710 (1995)].

\bibitem{OPAL:1996ikk}
{\bfseries OPAL} Collaboration, G.~Alexander {\em et~al.}, ``{A Study of charm hadron production in Z0 ---{\ensuremath{>}} c anti-c and Z0 ---{\ensuremath{>}} b anti-b decays at LEP},'' \href{https://dx.doi.org/10.1007/s002880050218}{{\em Z. Phys. C} {\bfseries 72} (1996) 1--16}.

\bibitem{ALICE:2023sgl}
{\bfseries ALICE} Collaboration, S.~Acharya {\em et~al.}, ``{Charm production and fragmentation fractions at midrapidity in pp collisions at $ \sqrt{\textrm{s}} $ = 13 TeV},'' \href{https://dx.doi.org/10.1007/JHEP12(2023)086}{{\em JHEP} {\bfseries 12} (2023) 086}, \href{https://arxiv.org/abs/2308.04877}{{\ttfamily arXiv:2308.04877 [hep-ex]}}.

\bibitem{CMS:2021lab}
{\bfseries CMS} Collaboration, A.~Tumasyan {\em et~al.}, ``{Measurement of prompt open-charm production cross sections in proton-proton collisions at $ \sqrt{s} $ = 13 TeV},'' \href{https://dx.doi.org/10.1007/JHEP11(2021)225}{{\em JHEP} {\bfseries 11} (2021) 225}, \href{https://arxiv.org/abs/2107.01476}{{\ttfamily arXiv:2107.01476 [hep-ex]}}.

\bibitem{Costantini:2024wby}
M.~N. Costantini, M.~Madigan, L.~Mantani, and J.~M. Moore, ``{A critical study of the Monte Carlo replica method},'' \href{https://dx.doi.org/10.1007/JHEP12(2024)064}{{\em JHEP} {\bfseries 12} (2024) 064}, \href{https://arxiv.org/abs/2404.10056}{{\ttfamily arXiv:2404.10056 [hep-ph]}}.

\bibitem{Kneesch:2007ey}
T.~Kneesch, B.~A. Kniehl, G.~Kramer, and I.~Schienbein, ``{Charmed-meson fragmentation functions with finite-mass corrections},'' \href{https://dx.doi.org/10.1016/j.nuclphysb.2008.02.015}{{\em Nucl. Phys. B} {\bfseries 799} (2008) 34--59}, \href{https://arxiv.org/abs/0712.0481}{{\ttfamily arXiv:0712.0481 [hep-ph]}}.

\bibitem{Salajegheh:2019nea}
M.~Salajegheh, S.~M. Moosavi~Nejad, M.~Soleymaninia, H.~Khanpour, and S.~Atashbar~Tehrani, ``{NNLO charmed-meson fragmentation functions and their uncertainties in the presence of meson mass corrections},'' \href{https://dx.doi.org/10.1140/epjc/s10052-019-7521-x}{{\em Eur. Phys. J. C} {\bfseries 79} no.~12, (2019) 999}, \href{https://arxiv.org/abs/1904.09832}{{\ttfamily arXiv:1904.09832 [hep-ph]}}.

\bibitem{Buckley:2014ana}
A.~Buckley, J.~Ferrando, S.~Lloyd, K.~Nordstr{\"o}m, B.~Page, M.~R{\"u}fenacht, M.~Sch{\"o}nherr, and G.~Watt, ``{LHAPDF6: parton density access in the LHC precision era},'' \href{https://dx.doi.org/10.1140/epjc/s10052-015-3318-8}{{\em Eur. Phys. J. C} {\bfseries 75} (2015) 132}, \href{https://arxiv.org/abs/1412.7420}{{\ttfamily arXiv:1412.7420 [hep-ph]}}.

\end{thebibliography}\endgroup

\end{document}